\documentclass[preprint, a4paper, 11pt, 3p, oneside, number, english]{elsarticle}
\usepackage[T1]{fontenc}
\usepackage[latin9]{inputenc}

\journal{SoftwareX}

\usepackage[pdftex]{color}
\usepackage{lineno}
\usepackage{babel}
\usepackage{amssymb}
\usepackage[font=footnotesize, labelfont=bf]{caption}
\usepackage[font=footnotesize, labelfont=bf]{subcaption}
\usepackage[hyphens]{url}
\usepackage[colorlinks=true,linkcolor=black, citecolor=blue, urlcolor=blue, unicode=true]{hyperref}


\begin{document}

\newcommand{\Angs}{\mathrm{\mbox{\AA}}}
\newcommand{\todo}[1]{\textbf{[TODO: #1]}}

% ------------------------------------------------------------------------------
% Front matter

\begin{frontmatter}{}
\title{Takin: An Open-Source Software for Experiment Planning, Visualisation, and Data Analysis}

\author[tum,mlz]{Tobias Weber \corref{cor}} \ead{tobias.weber@tum.de}
\author[tum,mlz]{Robert Georgii}
\author[tum]{Peter B\"oni}

\cortext[cor]{Corresponding author}
\address[tum]{Physik-Department E21, Technische Universit\"at M\"unchen, James-Franck-Str. 1, 85748 Garching, Germany}
\address[mlz]{Heinz-Maier-Leibnitz-Zentrum (MLZ),  Technische Universit\"at M\"unchen, Lichtenbergstr. 1, 85747 Garching, Germany}

\begin{abstract}
Due to the instrument's non-trivial resolution function, measurements on triple-axis spectrometers
require extra care from the experimenter in order to obtain optimal results and to
avoid unwanted spurious artefacts. We present a free and open-source software system
that aims to ease many of the tasks encountered during the planning phase, in the
execution and in data treatment of experiments performed on neutron triple-axis spectrometers.
The software is currently in use and has been successfully tested at the MLZ, but can be
configured to work with other triple-axis instruments and instrument control systems.

\vspace{0.25cm}

This was the 2016 pre-print of our software paper \cite{Takin2016}, which is now available
at \url{http://dx.doi.org/10.1016/j.softx.2016.06.002} and is licensed under CC BY 4.0
(\url{http://creativecommons.org/licenses/by/4.0/}). As pre-print, it is a modified form of the published paper.
Furthermore, small updates have been included in the present version, e.g. concerning the references.
\end{abstract}


\end{frontmatter}{}
% ------------------------------------------------------------------------------

%\linenumbers

% ------------------------------------------------------------------------------
%  (_)_ __ | |_ _ __ ___
%  | | '_ \| __| '__/ _ \
%  | | | | | |_| | | (_) |
%  |_|_| |_|\__|_|  \___/
% ------------------------------------------------------------------------------
\section{Motivation and significance}
\label{sec:intro}
The invention of the neutron triple-axis spectrometer by Brockhouse \cite{Brock1961}
has entailed a huge boost in the understanding of dynamics in solid matter.
Today, triple-axis spectroscopy has become a standard tool in the investigation
of materials in such diverse fields as magnetism, lattice dynamics, superconductivity,
and critical phenomena.

A small selection of very recent examples include an investigation of the interactions governing the
nematic order of iron-based superconductors \cite{Wang2016}, helimagnetic band formation
in the chiral magnet MnSi \cite{Max2015}, and the behaviour of the helimagnetic bands
at the conical-ferromagnetic phase transition \cite{ConiPaper}.
But despite their ubiquity in neutron physics, the triple-axis spectrometer
is often poorly understood by users in terms of its very complicated resolution
function. An understanding of the instrumental resolution is crucial for a correct
planning of measurements and data treatment, as resolution effects often lead
to artefacts in the spectra -- called ``spurions'' -- which cannot be easily
distinguished from genuine excitations.

In practice, resolution analysis has led to isolated per-instrument software
applications which are often either very complicated
to use, too specialised, or lacking the required features (e.g. \cite{Saroun97, rescal5, Boehm2013}).
In this article we present the free and open-source software system \textit{Takin},
which is capable of performing all the required steps in one package and which is
general enough to be easily adaptable to new triple-axis instruments. The
software furthermore features a fully-integrated graphical user interface (GUI)
and is therefore much more accessible than previous command-line tools.

%First we report about the feature sets of the \textit{Takin}
%software package and then we present one application of the software to the determination of
%the phonon spectrum in a spinel vanadate. \todo{mvo convo}

% ------------------------------------------------------------------------------
%      _                     _       _   _
%   __| | ___  ___  ___ _ __(_)_ __ | |_(_) ___  _ __
%  / _` |/ _ \/ __|/ __| '__| | '_ \| __| |/ _ \| '_ \
% | (_| |  __/\__ \ (__| |  | | |_) | |_| | (_) | | | |
%  \__,_|\___||___/\___|_|  |_| .__/ \__|_|\___/|_| |_|
%                             |_|
% ------------------------------------------------------------------------------
\section{Software description}
\label{sec:desc}

\begin{figure*}[htb]
	\centering
	\includegraphics[width=0.87\textwidth]{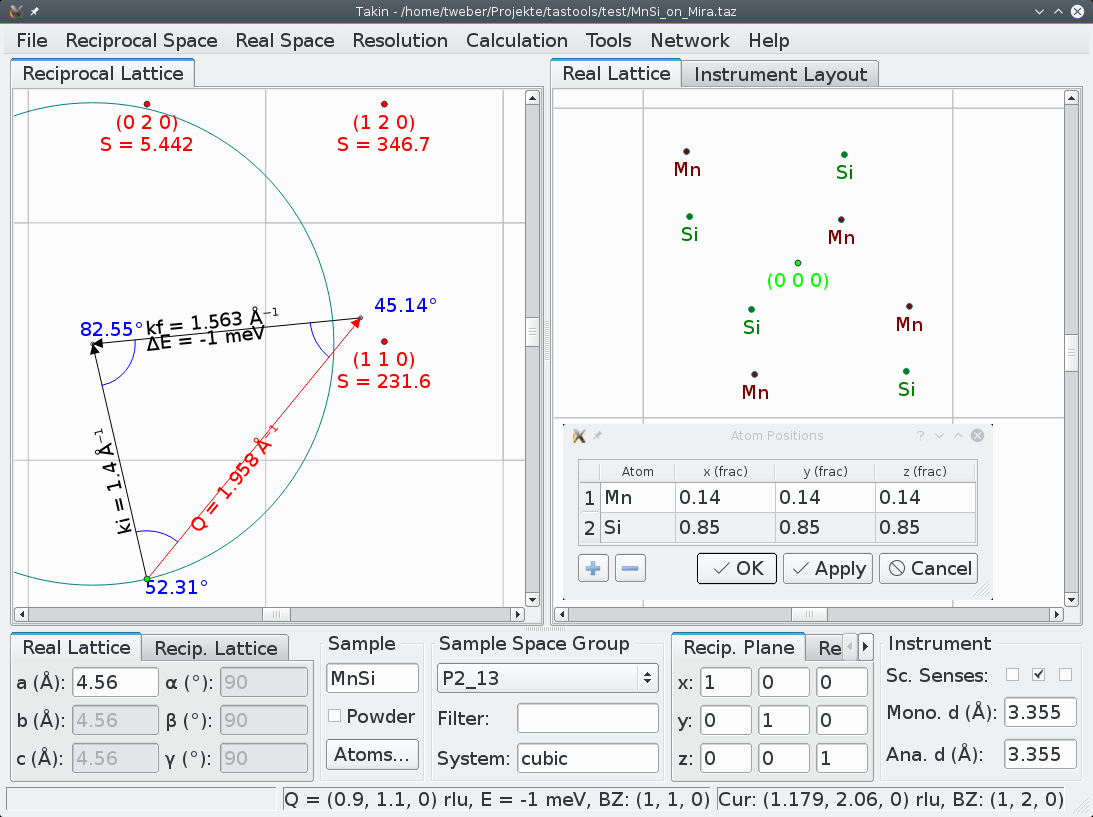}
	\caption{\textit{Takin} main view. Given the sample and instrument parameters, the
software displays the reciprocal lattice (left) and the real lattice (right).
In the reciprocal view, the neutron structure factors $S_{hkl}$ of the Bragg
peaks are shown and the first Brillouin zone is calculated, taking into account any elastic
peaks which are forbidden by the sample space group symmetry. The scattering triangle
can be dragged \& dropped and is translated into the corresponding triple-axis
instrument position (not shown) and resolution (see figure \ref{fig:mainres}).
The real space view shows the crystal lattice, the Wigner-Seitz cell and the
atomic positions in the selected unit cell. }
\label{fig:main}
\end{figure*}

\begin{figure}[ht!]
	\centering
	\includegraphics[width=0.91\columnwidth]{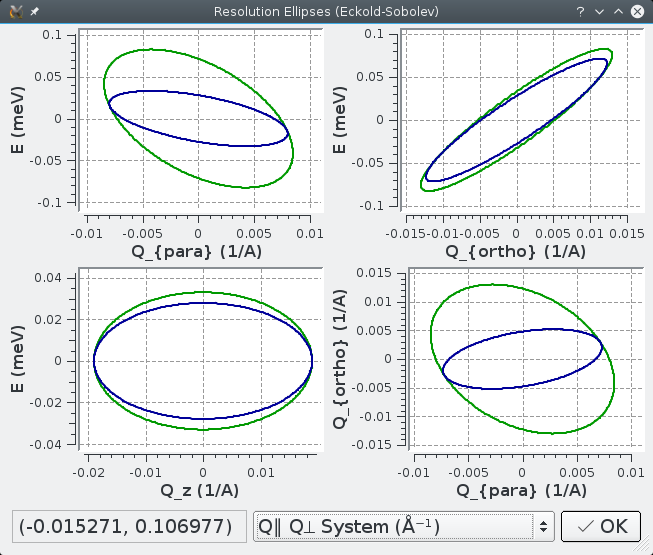}
	\caption{View of the corresponding resolution function of the triple-axis spectrometer.
	The software performs a realtime update and live display of the instrument resolution.
	Supported algorithms include the classical Cooper-Nathans and Popovici methods as well
	as the novel Eckold-Sobolev approach to the resolution function.}
\label{fig:mainres}
\end{figure}

\textit{Takin} is a software system encompassing multiple goals.
Its basic feature, namely a quick visualisation of both the crystal lattice in real
and in reciprocal space and the instrument configuration, is similar to the
\textit{vTAS} software \cite{Boehm2013}. But whereas \textit{vTAS} is mainly focused
on simulating realistic instrumental limits, e.g. walls in the instrument space,
\textit{Takin} does not include such features,
but instead focuses on combining crystallographic information and instrument
visualisation with a special emphasis on resolution calculation. Moreover,
a further emphasis of the software is data treatment and resolution deconvolution using
the Monte-Carlo approach.

% ------------------------------------------------------------------------------
%   __                  _   _                   _ _ _   _
%  / _|_   _ _ __   ___| |_(_) ___  _ __   __ _| (_) |_(_) ___  ___
% | |_| | | | '_ \ / __| __| |/ _ \| '_ \ / _` | | | __| |/ _ \/ __|
% |  _| |_| | | | | (__| |_| | (_) | | | | (_| | | | |_| |  __/\__ \
% |_|  \__,_|_| |_|\___|\__|_|\___/|_| |_|\__,_|_|_|\__|_|\___||___/
% ------------------------------------------------------------------------------
\subsection{Software Functionalities}
\label{sec:func}

The main window of the program is divided into two views: a reciprocal and a real-space
representation. The reciprocal space includes the Bragg reflections that are allowed
by the space group's symmetry operations as well as the elastic structure factors $S_{hkl} = \left|F_{hkl}\right|^2$ for each peak.
Given the allowed peaks, the first Brillouin zone is calculated and displayed.
Furthermore, the scattering triangle is drawn in the reciprocal space and can be either moved
by mouse or imported from scan files (the supported formats include the ones used by MLZ,
PSI, ILL and NIST). Alternatively, the software can receive the current live instrument position
by a network connection to the triple-axis control software. With respect to network connections,
two instrument control systems are currently supported: \textit{NICOS} \cite{Nicos},
the networked instrument control system which is used at the MLZ,
and \textit{SICS} \cite{Heer97}, the SINQ instrument control software used at the PSI.
Moreover, the object-oriented network architecture in \textit{Takin} allows for an easy
integration of further triple-axis control systems. The second part of the screen is
reserved for the real-space view, which shows the real crystal lattice and the Wigner-Seitz
cell. The unit cell and its atom positions, which are calculated by the given space group
symmetry operations, are shown in a projected representation in this view. As with the
reciprocal view, the displayed crystal plane can be freely chosen.

The software warns if the spectrometer -- either the real or the simulated one -- is
in a spurious scattering configuration. To that end \textit{Takin} includes a module
to check for the most important false peak conditions as given by Shirane \textit{et al.} \cite[Ch. 6]{Shirane2002}.
These include the Currat-Axe spurion which -- for example -- appears if $\mathbf{q}$
is parallel to $\mathbf{k_f}$ and $\left| \mathbf{k_f} + \mathbf{q} \right| = \left|\mathbf{k_i}\right|$.
In this configuration the analyser will Bragg-scatter an elastic incoherent signal from
the sample which subsequently appears as a faux-inelastic peak. Other spurious
signals that are checked for include the higher-order reflections from the monochromator
and the analyser and the Bragg tail. The latter appears in the $\mathbf{q}$ direction transverse
to the lattice vector $\mathbf{G}$
and results from the sloping of the resolution ellipse with respect to the
direction perpendicular to the mean $\mathbf{Q}$ and $E$ position.
Furthermore, a module for calculating powder peaks is included for a quick
assessment of spurious powder scattering from the sample environment, for instance
from aluminium or copper.
Please note that more specific spurious signals -- such as thermal diffuse scattering
by the analyser -- are not considered in the software.

The entire software package and its associated library \cite{tlibs} are written in
modern C++11 \cite{Stroustrup2013} and make extensive usage of the
\textit{Boost} \cite{Boost} C++ template library. The graphical user interface is
based on \textit{Qt} \cite{Qt} and the \textit{Qwt} \cite{Qwt} plotting library.
Tables with physical information, especially the properties of the 230 space group
types and the scattering lengths, are obtained from the \textit{Clipper} \cite{Clipper}
library and online from the NIST tables \cite{NISTscatlens}, respectively. We use
the magnetic form factors from the ILL tables \cite{ILLformfacts}.

\subsubsection{Resolution}
\paragraph{Resolution Calculation}
The basic resolution calculation module features a C++ re-implementation
of the Cooper-Nathans \cite{Cooper67} and Popovici \cite{Popovici75} algorithms
from Rescal 5 \cite{rescal5}. Furthermore, a novel implementation of the Eckold-Sobolev
approach to the resolution function \cite{Eckold2014} is available.
The latter algorithm is especially interesting for a realistic treatment of the common feature
of monochromator and analyser focusing, which increases the neutron flux, but also
increases the resolution volume. Each algorithm calculates the covariance
matrix for the wavevector $\mathbf{Q}$
and energy transfer $E$ of the neutron at a specific instrument position. The inverse
of the covariance matrix is the instrumental resolution matrix $R$ -- a quadric
describing a four-dimensional ellipsoid. The angles of the ellipsoid with respect
to the $\mathbf{Q}$ and $E$ axes and the widths with respect to the ellipsoid
axes are determined using the principal axis theorem
which involves solving the eigenvector problem. The software uses a specialised,
fast implementation for this reduced eigenvector calculation involving symmetrical
matrices. The result is displayed (see figure \ref{fig:mainres}) using
two- and three-dimensional projections of the ellipsoid in either
the $\left< \mathbf{Q_{\parallel}}, \mathbf{Q_{\perp}}, \mathbf{Q_z}, E \right>$
or the fractional $\left<h, k, l \right>$ crystal coordinate system.

The native C++ implementation of the algorithms is fast enough for a real-time updating
and displaying of the resolution function even on low-end hardware. On multi-core
or multi-processor systems further performance is gained by taking advantage of
the symmetric treatment of the monochromator and the analyser resolution volumes
in the Eckold-Sobolev method which allows for an efficient parallelisation of
the calculations.

\paragraph{Resolution Convolution}
Measurements using a triple-axis spectrometer do not yield the actual dynamical structure
factor $S\left(\mathbf{Q},\omega\right)$ (where $\omega = E/\hbar$), but instead give a convolution of $S$ with the
instrumental resolution $R$. Using an externally supplied theoretical $S\left(\mathbf{Q},\omega\right)$
function, an experiment can thus be simulated, e.g. for planning future measurements,
given the $R$ matrix.

\textit{Takin} contains a module for Monte-Carlo simulation of the convolution
integral (see figure \ref{fig:convo}). The Monte-Carlo approach includes generating
a predefined number of neutrons inside the four-dimensional resolution ellipsoid,
followed by sampling $S\left(\mathbf{Q},\omega\right)$ with the $\mathbf{Q}$ and $E$
coordinates of the Monte-Carlo neutrons.

The $S\left(\mathbf{Q},\omega\right)$ function for the excitations to be simulated can
be supplied as either a table or as a Python \cite{vanRossum2011} script or -- in case
performance is crucial -- as a native C++ class which has to be derived from a special
interface. For tabulated $S\left(\mathbf{Q},\omega\right)$ values, the values from the table are
used to generate a four-dimensional binary search tree (a k-d tree \cite{Bentley75}) which can then
be very efficiently used to query nearest positions by the Monte-Carlo simulation.
The Monte-Carlo simulation itself is sped up by making use of all available processor
cores for all model sources except the Python scripts (the latter being due to threading
limitations in the Python interpreter).

\paragraph{Resolution Convolution Fit}
Free parameters can be defined in the different theoretical $S\left(\mathbf{Q},\omega\right)$
models. In the Python model -- for instance -- all global variables are automatically determined
upon loading the script and offered as fit parameters in the convolution fitter tool.
User-selected free parameters are used to perform a $\chi^2$ minimisation in order to
get the optimal fit of the resolution convolution to experimental measurements. For performing the
minimisation, the Minuit software library \cite{Root2011} is used.

Due to its complexity, the fitter program module is currently only accessible via a
command-line interface and takes a special job file format as input. In order to be in
unison with the otherwise fully GUI-based design of the software a fully graphical convolution
fitter is planned for the near future.

% ------------------------------------------------------------------------------
%                                 _
%   _____  ____ _ _ __ ___  _ __ | | ___
%  / _ \ \/ / _` | '_ ` _ \| '_ \| |/ _ \
% |  __/>  < (_| | | | | | | |_) | |  __/
%  \___/_/\_\__,_|_| |_| |_| .__/|_|\___|
%                          |_|
% ------------------------------------------------------------------------------
\section{Illustrative Example}
\label{sec:ex}
\begin{figure}[ht!]
	\centering
% IMPORTANT: Use width=0.87\columnwidth for the final two-column version!
%	\includegraphics[width=0.87\columnwidth]{figures/convo}
%	\includegraphics[width=0.87\columnwidth]{figures/convo2}
	\includegraphics[width=0.36\paperwidth]{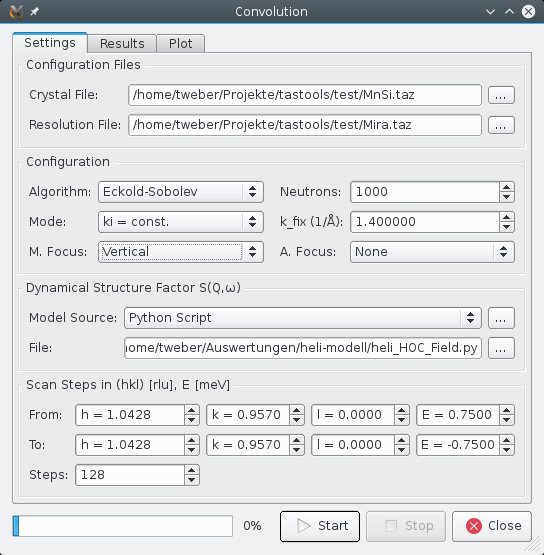}
	\includegraphics[width=0.36\paperwidth]{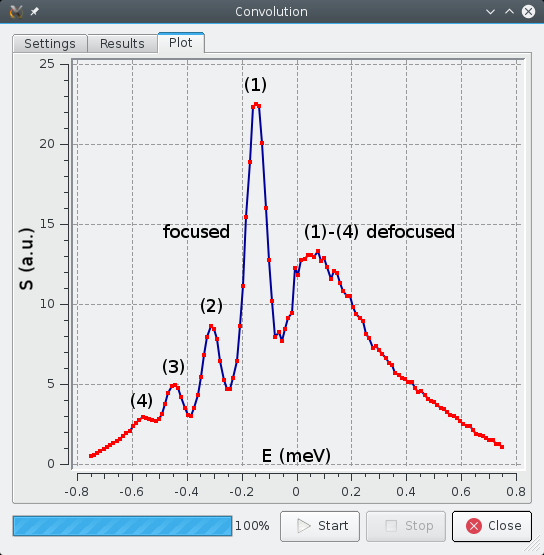}
	\caption{\textbf{Upper panel}: For experiment planning, a Monte-Carlo integration of the resolution convolution
	integral can be quickly done by selecting a previously defined crystal, instrument and theoretical
	$S\left(\mathbf{Q},\omega\right)$ model. A scan is then simulated along the defined path.
	\textbf{Lower panel}: Shown are the results of the scan defined in the upper panel,
	i.e. an energy scan at $\mathbf{Q} = \left( 1.043\ 0.957\ 0 \right)$.
	This example plot depicts a 4D convolution of the MIRA instrumental function with the
	helimagnons in a MnSi single crystal using the theoretical model developed by J. Waizner and M. Garst
	and implemented by G. Brandl \cite{Max2015}. Four helimagnon bands can be resolved in the focusing
	direction, which for $k_i = const.$ and the given $\mathbf{Q}$ is at $E < 0$. In the
	non-focusing direction at $E > 0$ the bands cannot be resolved and appear as one broad peak.
	The incoherent elastic contribution is not visible. }
	\label{fig:convo}
\end{figure}

As \textit{Takin} contains many different triple-axis features, we restrict
ourselves to showing one specific functionality here: planning an experiment using a
simulated scan, which can be fully and quickly done in the GUI. In the convolution dialog
(figure \ref{fig:convo}, upper panel), a file describing the crystal and the
instrument geometry have to be given along with a model file containing the
dynamical structure factor $S\left(\mathbf{Q}, \omega \right)$.
The crystal and instrument files are defined beforehand in the \textit{Takin} main view
(figure \ref{fig:main}).

As described earlier, the $S\left(\mathbf{Q}, \omega \right)$ model file can be either a C++ class,
a table or a Python script. In the present example, we use the chiral magnet MnSi as a
sample, MIRA \cite{MIRA, MIRAnew} as the instrument and the helimagnon model from \cite{Max2015} as a theoretical
$S\left(\mathbf{Q}, \omega \right)$ function. An arbitrary scan path in $\left( \mathbf{Q}, E \right)$
space can next be defined and simulated. Here, we simulate an energy scan at
$\mathbf{Q} = \left( 1.043\ 0.957\ 0 \right) \mathrm{rlu}$ using a fixed incident wavenumber,
$k_i = 1.4\,\Angs^{-1}$.
The resulting plot (figure \ref{fig:convo}, lower panel) shows that scans at this $k_i$
are viable on MIRA as four individual helimagnon bands can be resolved for $E < 0$,
which is the focusing direction for the defined positions. For the defocused measurement
positions at $E>0$ the simulation shows that the bands cannot be resolved anymore and
instead only one broad, smeared-out peak will be discernible.

\begin{figure}[ht!]
	\centering
	\includegraphics[width=0.75\columnwidth]{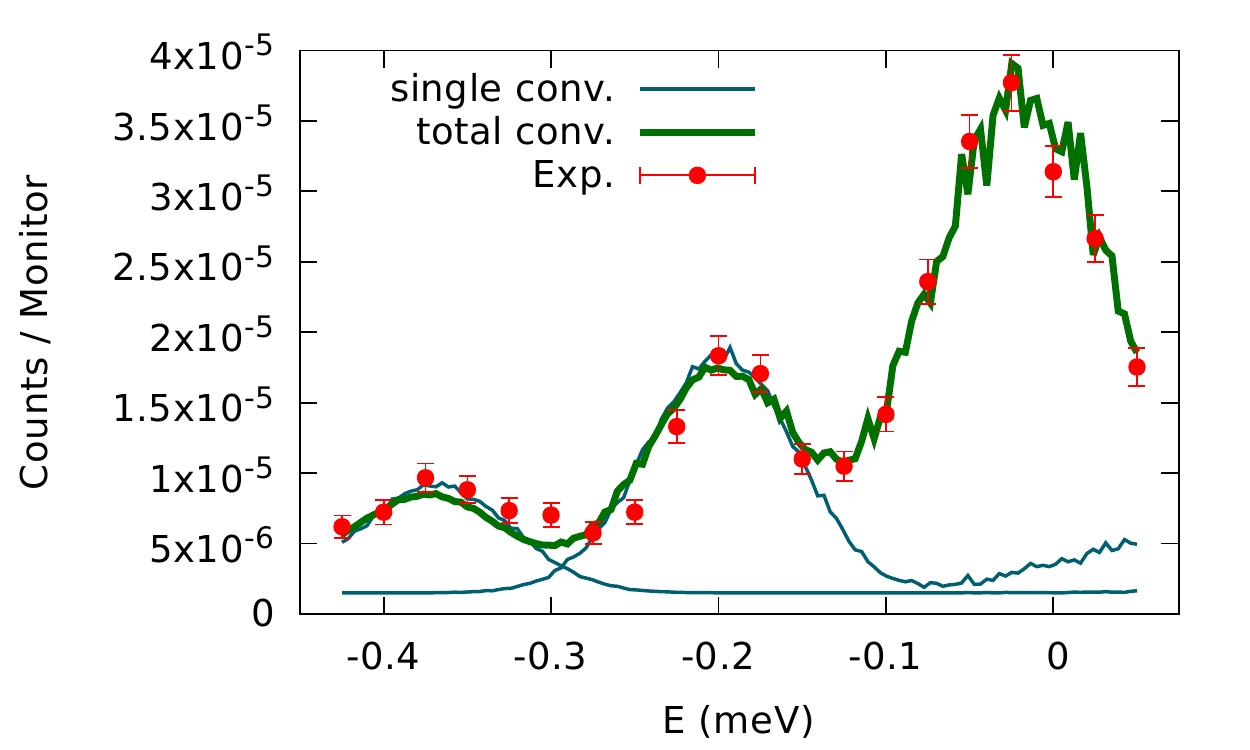}
	\caption{An example convolution fit to experimental data \cite{ConiPaper}.
	The data points were measured on MIRA \cite{MIRA, MIRAnew}, the lines are obtained from
	the \textit{Takin} convolution fitter using the model given in \cite{Max2015}.
	The momentum transfer $q$ of this data set differs from the one shown in figure
	\ref{fig:convo}.}
	\label{fig:convofit}
\end{figure}

In a further step, \textit{Takin} can be used for convolution fits to experimentally
measured data. We can confirm an excellent agreement of the calculated resolution
functions using a focusing monochromator and measurements of the helimagnon bands in
MnSi \cite{ConiPaper} as depicted in figure \ref{fig:convofit}.

% ------------------------------------------------------------------------------
%  _                            _
% (_)_ __ ___  _ __   __ _  ___| |_
% | | '_ ` _ \| '_ \ / _` |/ __| __|
% | | | | | | | |_) | (_| | (__| |_
% |_|_| |_| |_| .__/ \__,_|\___|\__|
%             |_|
% ------------------------------------------------------------------------------
\section{Impact}
\label{sec:impact}
With \textit{Takin} we hope to ease the life of triple-axis users in all
aspects of experiment planning, execution and data analysis. In the planning phase
it can be used to quickly and graphically assess if the experiment is
possible on the given instrument: for example if the angles are in the allowed range;
whether the scattering triangle is closed; or the resolution is good enough.

During the experiment, the software shows the live position
of the instrument in reciprocal space, draws the corresponding resolution function
at this position and warns of possible spurions. After the experiment, \textit{Takin}
can be used ``offline'' to do convolution simulations and convolution fits to the
measured spectra providing a choice between different resolution algorithms.

The latest version of the software is currently in regular use at the
MLZ instruments MIRA \cite{MIRA, MIRAnew} and PANDA \cite{PANDA}. Moreover, the software has
been successfully tested to work at the SINQ instrument TASP \cite{Semadeni01}
in ``live mode'' during a magnon measurement.

The first comprehensive applications for the resolution fitting capabilities of \textit{Takin}
will be presented in two upcoming papers by the authors. The first one is concerned with
the evolution of the helimagnon band structure in the vicinity of the conical-ferromagnetic
phase transition of MnSi \cite{ConiPaper}, the second one examines an anomalous phonon
behaviour at the cubic-tetragonal transition in the spinel vanadate MgV$_2$O$_4$ \cite{SpinelPaper}.

% ------------------------------------------------------------------------------
%                       _           _
%   ___ ___  _ __   ___| |_   _ ___(_) ___  _ __
%  / __/ _ \| '_ \ / __| | | | / __| |/ _ \| '_ \
% | (_| (_) | | | | (__| | |_| \__ \ | (_) | | | |
%  \___\___/|_| |_|\___|_|\__,_|___/_|\___/|_| |_|
% ------------------------------------------------------------------------------
\section{Conclusion}
\label{sec:conc}
We presented a software system for performing calculations and visualisations of
inelastic neutron scattering with a special emphasis on the resolution function of
triple-axis spectrometers.

For future developments, it is planned to offer general support for time-of-flight
spectrometers. Currently, the software already contains a module for the resolution
calculation and convolution using the Violini method \cite{Violini2014}.
Furthermore the convolution fitter will be equipped with advanced scripting
capabilities and also included in the main GUI.

% ------------------------------------------------------------------------------
\section*{Acknowledgements}
This work was supported by the German Research Foundation (DFG) and the Technische
Universit\"at M\"unchen within the funding programme Open Access Publishing.
The work at MIRA was supported by the DFG under GE971/5-1.

We wish to thank Max Kugler and Georg Brandl for many discussions about triple-axis
spectroscopy and instrumentation, and for providing the helimagnon Python
script for the convolution example.
We also want to thank Petr \v{C}erm\'ak, Astrid Schneidewind, and Bertrand Roessli
for their support and for providing PANDA and EIGER resolution parameters,
respectively.

This work is based on experiments performed at the instrument MIRA operated by FRM II
at the Heinz Maier-Leibnitz Zentrum (MLZ), Garching, Germany.

The software's DOI is: \href{http://dx.doi.org/10.5281/zenodo.3961491}{10.5281/zenodo.3961491}.

% Warning: Forgot this section before.
\section*{Author Contributions}
T.W. designed and developed the software and originally wrote the manuscript as part of his Ph.D.
thesis \cite{PhDWeber}. P.B. and R.G. contributed to the manuscript and to the design of the software through
many discussions over the years.
T.W. conducted the experiment at MIRA, R.G. was the local contact, and P.B. the supervisor.

% ------------------------------------------------------------------------------

\section*{Required Metadata}
\label{}

\section*{Current code version}
\label{}

\begin{table}[!ht]
\begin{tabular}{|l|p{6.5cm}|p{6.5cm}|}
\hline \textbf{Nr.}	& \textbf{Code metadata description} 							&  \\
\hline C1 			& Current code version 											& 1.0 \\
\hline C2 			& Permanent link to code/repository used for this code version	& \url{https://forge.frm2.tum.de/cgit/cgit.cgi/frm2/mira/tastools.git/} and \url{https://github.com/t-weber/takin} \\
\hline C3 			& Legal Code License   											& GPL Version 2 \\
\hline C4 			& Code versioning system used  									& git \\
\hline C5 			& Software code languages, tools, and services used 			& C++ 11, CMake 3 \\
\hline C6 			& Compilation requirements, operating environments  			& Linux, OS X, Unix like, Windows (via MinGW); GCC 4.8 (or later) or Clang, Boost,
	Qt 4 or 5, Qwt 5 or 6, Clipper, Minuit 2 \\
\hline C7 			& If available Link to developer documentation/manual 			& \url{https://forge.frm2.tum.de/cgit/cgit.cgi/frm2/mira/tastools.git/plain/doc/index_help.html} \\
\hline C8 			& Support email for questions 									& tobias.weber@tum.de \\
\hline
\end{tabular}
\caption{Code metadata}
\label{}
\end{table}

\end{document}

% ------------------------------------------------------------------------------
% Front matter
% ------------------------------------------------------------------------------
\begin{frontmatter}{}
\title{Update 2.0 to ``Takin: An Open-Source Software for Experiment Planning, Visualisation, and Data Analysis'', (PII: S2352711016300152)}

\author[ill]{Tobias Weber \corref{cor}} \ead{tweber@ill.fr}

\cortext[cor]{Corresponding author, \url{https://orcid.org/0000-0002-7230-1932}}
\address[ill]{Institut Laue-Langevin (ILL), 71 avenue des Martyrs, 38000 Grenoble, France}

\begin{abstract}
We report a major new release of the inelastic neutron scattering software \textit{Takin}.
The software package comprises programme modules for experiment planning, scan visualisation and data treatment at neutron triple-axis spectrometers. 
The new version 2.0 presents enhancements of existing tools and introduces new modules.

\vspace{0.25cm}

This was the 2021 pre-print of our software paper \cite{Takin2021}, which is now available
at \url{https://dx.doi.org/10.1016/j.softx.2021.100667} and is licensed under CC BY 4.0
(\url{http://creativecommons.org/licenses/by/4.0/}). As pre-print, it is a modified form of the published paper.
Furthermore, small updates have been included in the present version.
\end{abstract}

\begin{keyword}
    triple-axis spectroscopy \sep
    resolution convolution fitting
\end{keyword}

\end{frontmatter}{}
% ------------------------------------------------------------------------------

\section{Introduction}
\label{sec:intro}

In the years since its previous releases, the software \textit{Takin} \cite{Takin, Takin2016, Takin2017} has been employed alongside experiments and during data treatment for a multitude of neutron triple-axis studies. The gained experience allowed us to continuously improve its modules, of which we present selected updates in this short paper.

\section{Resolution convolution simulator and fitter}
\label{sec:convofit}

\begin{figure*}[htb]
	\centering
	\includegraphics[width=1\textwidth]{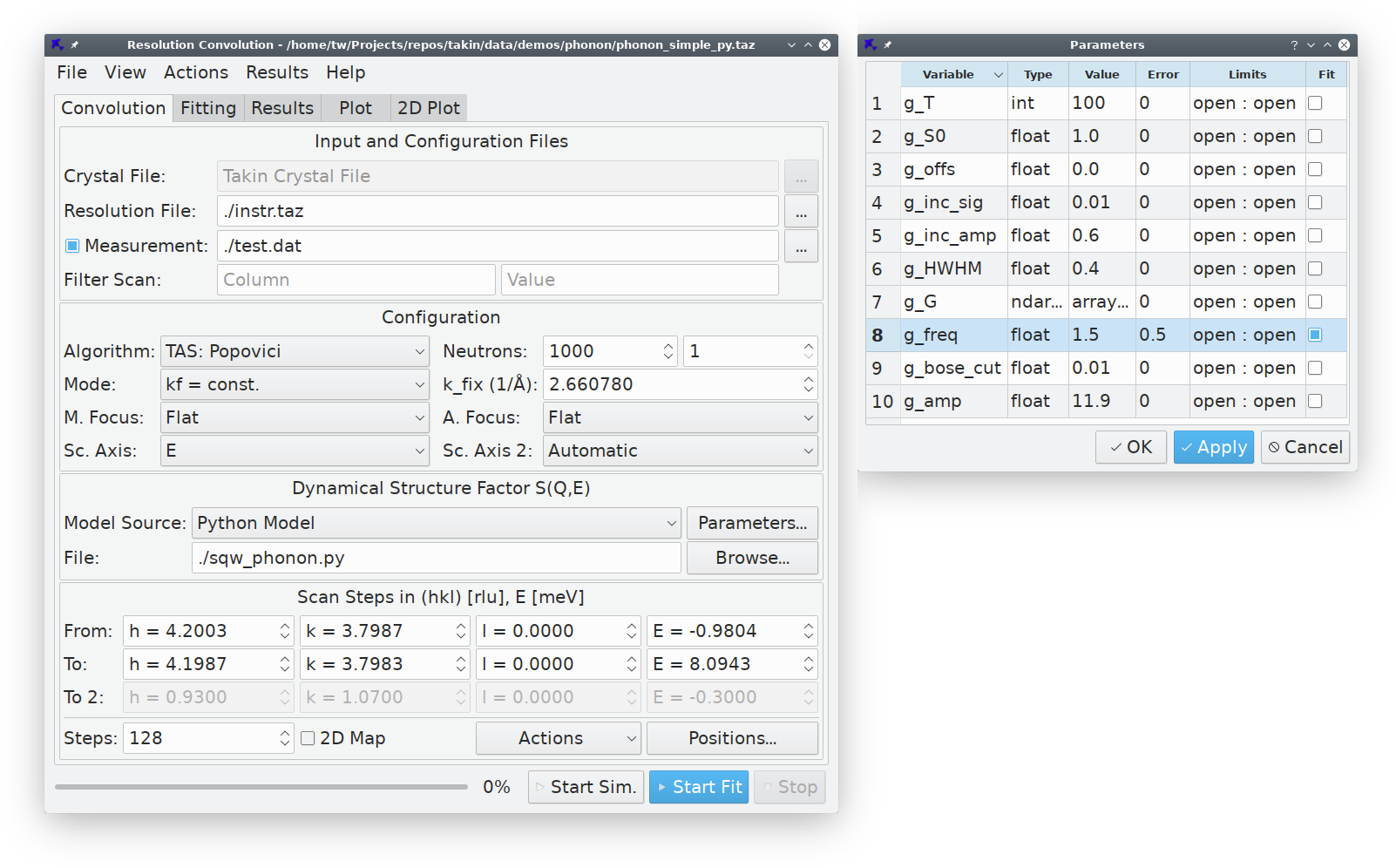}
	\caption{The convolution programme module (left panel) now allows to conveniently set up and perform resolution convolution fits directly from within its GUI. To that end, it suffices that the desired fit parameter be marked as such in the parameters dialog (right panel). Clicking ``Start Fit'' starts the fitting process.}
	\label{fig:convo}
\end{figure*}

The previous versions 1.0 and 1.5 of \textit{Takin} lacked a graphical user interface (GUI) for one of its central components, namely the resolution convolution fitter. Job files needed to be exported, adapted, and run using an external command-line programme. 
The new version now adds the missing GUI functionality. Having a graphical fitter greatly simplifies the complex task of setting up and performing resolution convolution fits: These can now be conveniently done from the basis of an existing simulation model with a simple click of a button, see Fig. \ref{fig:convo}. The fitter itself is based on ROOT/Minuit2 \cite{ROOT}.

The resolution convolution simulator had the opposite problem: It did feature a GUI module in the previous version, but lacked a command-line interface (CLI). This too has now been added, allowing for an easy set up of batch jobs.

The convolution fitter and simulator employ a plug-in system for interfacing with user-provided dynamical structure factor models. For the new version, the internal structure of the plug-in system has been reworked from scratch. The goal is to provide better stability by isolating the main programme and the plug-in processes. 

In \textit{Takin} 1.0, for instance, the \textit{Python} \cite{Python1995, Python2018} structure factor scripting module directly executed inside the main process space, severely limiting multi-threading capabilities. In \textit{Takin} 1.5, the module still ran in the main process, but spawned the \textit{Python} interpreters as sub-processes. This made multi-processing possible, but could still lead to stability problems if a script misbehaved. In the new version, all dependencies to \textit{Python} have been removed from the \textit{Takin} main process and a layer of indirection has been added. \textit{Takin} now spawns a \textit{Python} control process, which itself creates the actual \textit{Python} sub-processes, ensuring both multi-processing capabilities and stability for the main process.

\section{Further updates}
\label{sec:misc}

A non-exhaustive list of further updates and additions includes:

\begin{itemize}
\item Crystal parameters can now be imported into the main GUI from the crystal information file format (CIF). To that end we make use of the \textit{Gemmi} library \cite{GEMMI}.
\item Inaccessible instrument angles can be defined and are displayed in the reciprocal space view of the main window. This is used to exclude scattering angles which cannot be reached, for example due to the sample being mounted in a cryomagnet.
\item Reflectivity curves of the monochromator crystal can be taken into account for resolution calculation of instruments which operate in a $k_i = const$ mode, for instance MIRA \cite{MIRA}.
\item A spline-based pre-fitter has been added to the scan data viewer module, improving initial guesses for curve fitting.
\item A covariance calculation tool is now available to calculate the resolution function of Monte-Carlo neutron events. This tool also forms the basis for the new \textit{mcresplot} programme in the neutron ray-tracing software \textit{McStas} \cite{McStas}. It is based on \textit{McStas'} original \textit{mcresplot.pl} tool \cite{McStas}.
\item The implementation of G. Eckold's and O. Sobolev's resolution algorithm \cite{Eckold14} now supports calculation for vertical scattering in the $k_f$ axis, based on new calculations and descriptions provided by G. Eckold \cite{Eckold20}.
\end{itemize}

\section{Conclusion and outlook}
\label{sec:mag}
The updates of the present version are based on experimental experience and user interactions, and will improve user friendliness as well as stability and performance.

Future updates to \textit{Takin} will see a better integration of magnetic structure visualisation and data treatment into the main programme. Nevertheless, several tools concerning magnetic and polarised scattering are already present in the current version and are ready to be used and tested. 
So far, these include programme modules that are concerned with drawing magnetic structures, calculating magnetic structure factors and visualising in an interactive fashion the vectors of the Blume-Maleev equation \cite{Brown2006}, which is employed in polarised scattering. These new tools are based on descriptions and formulas from \cite{Chatterji2006, Schweizer2006, Brown2006, Brown2006_2}.

% ------------------------------------------------------------------------------
\section*{Acknowledgements}
The open access fee was covered by FILL2030, a European Union project within the European Commission's Horizon  2020 Research and Innovation programme under grant agreement No. 731096.
Many thanks go to Fr\'ed\'eric Bourdarot for his invaluable feedback on the software, bug reports and beta-testing of new features.
Many thanks also to Alain Filhol for preparing an App Store bundle for the MacOS version and for extensive testing and bug hunting.
Additionally, we wish to thank Markos Skoulatos, Lukas Beddrich, Marc de Boissieu, Petr \v{C}erm\'ak, Astrid Schneidewind, Jitae Park, Greg Tucker, Georg Brandl, Chris Franz and Bertrand Roessli for helpful feedback and bug reports.
Special thanks to Peter B\"oni and Robert Georgii, the author's former thesis supervisors during the first version of the software; as well as to Paulo Mutti for his support.
The software DOIs are \href{http://dx.doi.org/10.5281/zenodo.4117437}{10.5281/zenodo.4117437} for the current version and \href{http://dx.doi.org/10.5281/zenodo.3961491}{10.5281/zenodo.3961491} for the previous one.

% ------------------------------------------------------------------------------
\section*{Required Metadata}
\label{metadata}

\section*{Current code version}
\label{metadata:code}

\begin{table}[!ht]
\begin{tabular}{|l|p{6.5cm}|p{6.5cm}|}
\hline \textbf{Nr.}	& \textbf{Code metadata description} 		&  \\
\hline C1 	& Current code version					& 2.0 \\
\hline C2 	& Permanent link to code/repository used for this code version	& \url{https://github.com/t-weber/takin2} \\
\hline C3 	& Legal Code License					& GNU GPL Version 2 for the core software and GPL Version 3 for the new modules \\
\hline C4 	& Code versioning system used			& git \\
\hline C5 	& Software code languages, tools, and services used 	& C++11 for core tools, C++20 for extension modules, \href{https://cmake.org}{CMake 3} \\
\hline C6 	& Compilation requirements, operating environments  	& Linux, MacOS, MinGW; \href{https://gcc.gnu.org}{GCC 9} or \href{https://clang.llvm.org}{Clang 10}, \href{https://www.boost.org}{Boost 1.7x}, \href{https://qt.io}{Qt 5}, \href{http://qwt.sourceforge.net}{Qwt 6}, \href{http://www.ysbl.york.ac.uk/~cowtan/clipper}{Clipper}, \href{https://project-gemmi.github.io}{Gemmi}, \href{https://root.cern}{Minuit 2}, \href{https://freetype.org}{Freetype}, \href{https://www.python.org}{Python 3}, \href{https://numpy.org}{NumPy}, \href{https://www.scipy.org}{SciPy} \\
\hline C7 	& If available Link to developer documentation/manual 	& Available from within the software. \\
\hline C8 	& Support email for questions			& tweber@ill.fr \\
\hline
\end{tabular}
\caption{Code metadata}
\label{}
\end{table}

\section*{Current executable software version}
\label{metadata:binary}

\begin{table}[!ht]
\begin{tabular}{|l|p{6.5cm}|p{6.5cm}|}
\hline \textbf{Nr.}	& \textbf{(executable) Software metadata description}	&  \\
\hline S1 	& Current software version				& 2.0 \\
\hline S2 	& Permanent link to executables of this version		& \url{http://wiki.mlz-garching.de/takin} \\
\hline S3 	& Legal Code License					& GNU GPL Version 2 for the core software and GPL Version 3 for the new modules \\
\hline S4 	& Computing platform / Operating System		& Linux, MacOS, Windows (experimental, via MinGW) \\
\hline S5 	& Installation requirements \& dependencies	&  \\
\hline S6 	& If available Link to user manual			& Available from within the software. \\
\hline S7 	& Support email for questions				& tweber@ill.fr \\
\hline
\end{tabular}
\caption{Executable metadata}
\label{}
\end{table}